\documentclass[pr,aps,praps,amsbsy]{revtex4}
\usepackage{graphics}
\begin{document}

\title{\normalsize{\bf{PHOTOPRODUCTION OF ELECTRON-POSITRON PAIRS IN BENT
SINGLE  CRYSTALS
}}}

\author{Yu.~A.~Chesnokov, V.~A.~Maisheev \\
{\it{Institute for High Energy Physics, 142281, Protvino, Russia}}\\
D.~Bolognini, S.~Hasan, M.~Prest \\
{\it{Universit\`a dell'Insubria and INFN Milano Bicocca, Como, Italy}} \\
E.~Vallazza \\
{\it{INFN Trieste, Padriciano 99, Trieste, Italy}}}

\begin{abstract}
The process of photoproduction of electron-positron pairs in bent single crystals is considered 
in this paper. In particular, it is shown that  the probability of the process for $\gamma$-quanta 
with energies from 100 GeV on is significantly higher than the one in an amorphous medium.   
A possible scenario for the experimental validation of the process is discussed and the positive features 
of the 
photoproduction in bent crystals compared to straight ones are underlined from the point of view of 
possible applications. 
\end{abstract}

\maketitle
PACS number(s): 61.80.Cb, 29.27.-a, 29.30.-h

\section{Introduction}
In recent years several experiments studying different processes of charged particle interaction with 
bent single crystals were performed \cite{WS, WS1, AA}. In particular, the process of radiation emission of
ultrarelativistic electrons and positrons moving in the planar fields of a bent single crystal
was investigated. The experiments have shown a significant increase of the radiation energy losses
of electrons and positrons in conditions of volume reflection \cite{WS1,AA}. \\
The present paper considers the process of $e^\pm$-pair creation by high energy photons moving 
in a bent single crystal with a constant curvature. This study is based on the theory of coherent 
$e^{\pm}$ pair production \cite{TM,BKS} in straight single crystals.  \\
The photoproduction of $e^\pm$-pairs in a periodically deformed single crystal was studied in
\cite{MSP}. This study and the consequent results depend essentially on the periodic properties of the 
process and hence cannot be used directly for the case of a bent single crystal with a constant curvature.
  
Sec.~II and III are devoted to the mathematical description of the motion of charged and neutral particles in 
bent crystals and of the photoproduction process while sec.~IV describes the application of this model 
to two possible photon energies (120 and 1000~GeV). Sec.~V presents a possible experimental scenario 
to validate the model, discussing its implementation.

\section{Motion of charged and neutral particles in the bent planar fields of a single crystal}
The motion of ultrarelativistic charged particles in bent single crystals can be 
described with the help of the following equations \cite{MV}:
\begin{equation}
  E_0 \beta^2 v_r^2/(2c^2) +U(r)+ E_0 \beta^2 (R-r)/r= E=const,
\end{equation}
\begin{equation}
  dy/dt=v_y=const,
\end{equation}
\begin{equation}
  v_z = rd\phi/dt \approx  c \left (1-{1\over 2\gamma^2}-{(v^2_r + v_y^2)\over 2c^2} \right).
\end{equation}
These equations are valid in the cylindrical coordinate system ($r, \phi, y$). 
$v_r$ is the component of the particle velocity in the radial direction, $v_y$ the one 
along the $y$-axis and $v_z$ the tangential one; 
$R$ is the bending radius of the single crystal, $E_0$  and $\gamma$ are the particle energy and
its Lorentz factor, $E$ is the constant value of the radial energy, $U(r)$ is the crystal one 
dimensional potential, $c$ the velocity of light and $\beta$ the ratio of the particle velocity
to the velocity of light. \\
This paper considers the planar case where the scattering 
is due to the interaction of the particles with the set of crystallographic planes located normally to 
the ($r, \phi$)-plane. 
In practice it means that $v_y/c \gg \theta_{ac}$ but $v_y/c \ll 1$ for ultrarelativistic particles,
where $\theta_{ac}$ is the critical angle of  axial channeling. 

\begin{figure}
\scalebox{0.7}{\includegraphics{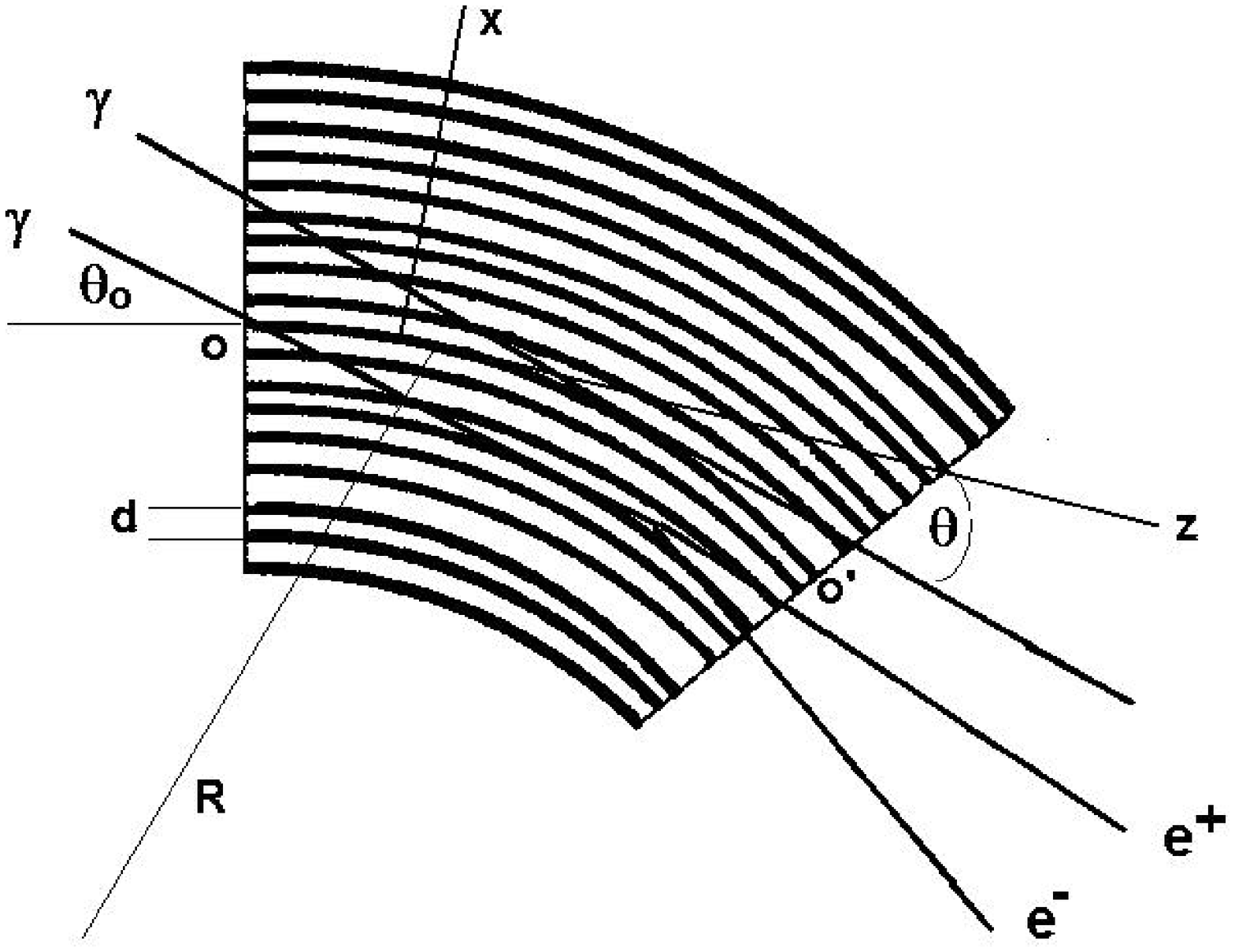}}
{\caption{}
\label{fig-1}}
\end{figure}

Eq.~(1) can be transformed in the following way:
\begin{equation}
  E_0 \beta^2 v_x^2/(2c^2) + U(x) +E_0\beta^2 x/R=E,
\end{equation}
where $x$ is the local Cartesian coordinate which is related to the cylindrical coordinate $r$
through the expression $x= R-r$ and $v_x=v_r$, and the $r$-value in the denominator of Eq.~(1) has been changed to $R$. 
This change introduces a negligible error in the experimental measurement (of the order of $x/R$).
Finally, $E$ is the transversal energy. \\ 
Fig.~\ref{fig-1} illustrates the geometry  of the $e^\pm$-pair production process and the local coordinate system
which will be used in the following considerations. \\
For neutral particles ($\beta=1$ for  photons) Eq.~(4) should be substituted with:
\begin{equation}
  {E_\gamma v_x^2 \over 2c^2} + {E_\gamma x \over R} = E,
\end{equation}
where $E_\gamma$ is the photon energy. This leads to: 
\begin{equation}
  x = x_0 +v_ot -{t^2 c^2 \over 2R},
\end{equation}
\begin{equation}
  \theta = {1\over c}{dx \over dt}= \theta_0 -{ct \over R},
\end{equation}
where $x_0$ and $v_o$ are the initial coordinate and velocity of the photon in the local Cartesian coordinate system 
($t=0$ corresponds to the $x_0$-coordinate and $v_o= c\theta_0$, and hence $E=E_\gamma v_0^2/(2c^2) + E_\gamma x_0/R$).
Eq.~(7) describes the variation of  the direction of the photon motion relative to the planar electric field 
(in the local coordinate system) and plays an important role in the description of the process. \\ 
 
The process of photoproduction in a bent single crystal is very similar to the process of radiation emission of electrons and
positrons undergoing volume reflection in such a crystal. The photoproduction probability 
depends on the character of the electron (positron) motion.
% in the place of location of the moving high energy photon. ???????
However, differently from radiation emission, photoproduction is a threshold process, which introduces 
some peculiarities. \\
The motion of charged particles was investigated in \cite{MV} and those results will be used in the following.

\section{Probability of photoproduction of $e^\pm$-pairs in the bent planes of a single crystal}

Let us consider the motion of a high energy photon which forms an initial entrance angle $\theta_0$ with the crystal 
and moves towards the tangency point. During its motion, the photon intersects the bent planes of the crystal 
(see fig.~\ref{fig-1}). 
In the local Cartesian coordinates, this motion is described by Eq.~(6).
At a time $t_1$ the photon has the corresponding coordinate $x_0 <x < x_c$, where $x_c$ is the 
coordinate of the tangency point. If it is assumed that the bending radius of the single crystal is 
large enough, the influence of the curvature on the process in some area $x_0<x_1 < x < x_2<x_c$ can be neglected 
allowing to use the theory of coherent pair production in straight crystals for the description of the process
in the area $x_1, x_2$. The theory of coherent pair production in crystals requires the following 
conditions to be respected \cite{TM,BKS}:
\begin{enumerate}
  \item the mean angle of particle deflection from the straight line on the formation length $l_f$ should be 
    $\ll 1/\gamma_f$, where $\gamma_f = E_\gamma/mc^2 $ ($m$ is the mass of the electron)
  \item the differences of the periods for neighbouring crossings should be small
  \item the characteristic parameter $\rho$ should be $\ll 1$
\end{enumerate}
The first condition in the case under consideration means that $\l_f/R \ll 1/\gamma_f$. 
The formation length for the process may be written as
\begin{equation}
  l_f^{-1} = {mc^2 \over 2 E_\gamma \lambda_c x_e(1-x_e)} \le {2 m c^2 \over E_\gamma \lambda_c},
\end{equation}
where $x_e$ is the relation of the electron (or positron) energy to the photon energy and $\lambda_c$ is 
the Compton wavelength of the electron.\\
It follows that: 
\begin{equation}
  R \gg \gamma^2 \lambda_c /2.
\end{equation} 
The second condition can be explained in the following way. In a straight single crystal the coherent process
of pair production takes place because of the periodical perturbation of the photon state at the crossing of 
the atomic planes. 
The same explanation can be also applied to a bent crystal in the limited area $x_1, x_2$. 
Finding $t$ as a function of $x$ (see Eq.~(6)) and taking into account that the atomic planes have a period 
equal to $d$, it is possible to calculate the time periods of the planes crossing during the photon motion. 
The total number of periods (from the point of entrance to the tangency point) is $\approx \theta^2 R/(2d)$. \\
With the help of Eqs.~(1)-(4), one can compute the parameter 
$\rho = 2 \gamma^2 \langle (V_x- \langle V_x \rangle)^2 \rangle/c^2$ for positrons or electrons.
Thus the following approximate relation can be written:
\begin{equation}
  \langle (V_x- \langle V_x \rangle)^2 \rangle \approx
          {c^2 \over 8E_0\beta^2} {(\langle U^2(x) \rangle- \langle U(x) \rangle^2)\over (E-E_0\beta^2 x/R)},
\end{equation}
where $\langle U(x) \rangle$ and $\langle U^2(x) \rangle$ are the mean and mean square of the planar potential
and $E\approx E_0\theta^2_c/2 +U(x_1)$, where the $\theta_c$ angle corresponds to the $x_1$ coordinate of 
the point of the particle creation.   
This relation is violated in the region close to the tangency point; in this region in fact a more exact 
calculation gives an approximately 
twice larger result. With the condition $E_0\beta^2 x/R \ll E$, it follows that: 
\begin{equation}
  \rho_e \approx {(\langle U^2(x) \rangle- \langle U(x) \rangle^2) \over 2m^2 c^4 \theta_c^2}.
\end{equation} 
Note that $\langle U^2(x) \rangle- \langle U(x) \rangle^2 = 42.3$~eV$^2$ for the silicon (110) plane. 

It should be noted that for coherent pair production, the following limit on the angle holds:
\begin{equation}
  \theta_l(n) \ge 2mc^2 /(E_\gamma \lambda_c Gn),
\end{equation}
where $n$ is a positive integer number and $G=2\pi/d$. $n$ is the number of the harmonics of the potential. 
It is well known that the  main contribution to the process is given by the first harmonic, 
while the contribution of the 4-th and higher order harmonics is small. Thus, Eq.~(12) shows that 
for small enough angles, the coherent process of pair production is absent.   
 
At this point, the conditions for coherent pair production in the (110) bent silicon planes can be investigated. 
The analysis will be performed for two photon energies equal to 120 and 1000~GeV. \\
From Eq.~(9) it follows that the bending radius has to be $R \gg 0.8$ and  $80$ cm respectively.
With the help of Eq.~(12) it is found that $\theta_l(1) $=0.4 and 0.04~mrad respectively, while  
the $\rho_e$ parameter should be  $\rho_e \le 1$ at $\theta_c \approx 0.01$~mrad.   \\
The  ratio of the neighbouring periods for R=250~cm at $\theta \approx 0.05$~mrad differs from 1 to 3$\%$,
while for R=1000~cm it differs of 1$\%$. Increasing the $\theta$-angle, this value becomes significantly better.  

It is evident that also in bent crystals the conditions for coherent pair production can be met in small areas: 
\begin{equation}
  d{\cal{W}}(L,\theta_0) = W(\theta(t))cdt,
\end{equation}
where ${\cal{W}}(L, \theta_0)$ is the total probability of pair production on the full length $L$ of a
single crystal by a photon with an entrance angle $\theta_0$ and $W(\theta)$ is the probability of pair production
per unit length by a photon moving with an angle $\theta$. Taking into account Eqs.~(6) and (7) it is found that: 
\begin{equation}
  {\cal{W}}(L,\theta_0)=R| \int^{\theta_0}_{\theta_1} W(\theta)d\theta |,
\end{equation}
where $\theta_1= \theta_0 - L/R $  in accordance with Eq.~(7) and the $W$-functions are equal to 
(see for example \cite{TM,BKS}): 
\begin{equation}
  W_0(\theta) = {W_{||}+ W_{\perp} \over 2}= 
  {nB\sigma_0 E_\gamma  \lambda_c\over 4 mc^2}
  \sum_{i=1}^\infty \Phi(Gi) (Gi)^2 F_2(z_i) + n\sigma_A,
\end{equation}
\begin{equation}
  D(\theta)={W_{||}-W_{\perp} \over 2}=
  {nB\sigma_0 E_\gamma \lambda_c \over 16 mc^2}
\sum_{i=1}^\infty \Phi(Gi)(Gi)^2 F_1(z_i),
\end{equation}
where  $\sigma_A$ is the cross section of pair production in a non oriented crystal  \cite{TM},
$\sigma_0=\alpha_e Z^2 r_e^2$, $\alpha_e$ is the fine-structure constant, $Z$ is the
atomic number of the crystal material, $r_e$ is the classical electron radius,
$B=16\pi^2/(N_S \Delta)$, $N_S$ is the number of atoms per elementary cell (with a volume $\Delta$) of a crystal,
$n=N_S/\Delta$ is the atomic density, $G=2\pi /d$ and $d$ is the interplanar distance.  
The variable $z_i$ is defined from the relation:
\begin{equation}
  z_i= {2 mc^2 \over E_\gamma \lambda_c Gi|\theta |}.
\end{equation}
Other definitios are:  
\begin{equation}
  \Phi(Gi)=|S(Gi)|^2(1-F(Gi))^2 e^{-A(Gi)^2}/(Gi)^4,
\end{equation}
where $S(Gi)$ is the structure factor of the reciprocal vector $Gi$, $F$ is the atomic form factor,
$A$ is the mean square of the thermal atomic vibrations. The functions $F_1,\,F_2$ are the following:
\begin{equation}
  F_1(z)=z^4(\ln{1+\sqrt{1-z}\over 1-\sqrt{1-z}}+ {2\sqrt{1-z}\over z})\eta(z)\eta(1-z),
\end{equation} 
\begin{equation}
  F_2(z)= z^2((1+z-z^2/2)\ln{1+\sqrt{1-z}\over 1-\sqrt{1-z}} -\sqrt{1-z}(1+z))\eta(z)\eta(1-z),
\end{equation}  
where the $\eta(z)$-function is equal to 1 or 0 at $z \ge 0$ and $z <0$ respectively. \\   
The equations presented here reflect the fact that the crystal can feel the linear polarization
of the photons and hence $W_{||}$ and $W_{\perp}$ are the probabilities of pair production per unit length for
photons with a linear polarization along a planar electric field and along a perpendicular plane 
respectively. Substituting these probabilities in Eq.~(14), one can obtain the total probabilities in a bent single crystal
for the corresponding photon polarization states. It is clear that the $W_0$-probability corresponds to
a unpolarized beam. After the integration (in accordance with Eq.~(14)), one has:
\begin{equation}
  {\cal{D}}_0 ={{\cal{W}}_{||}-{\cal{W}}_\perp \over 2} ={nB\sigma_0 R \over 8} 
  \sum_{i=1}^\infty \Phi(Gi)Gi |{\cal{F}}_1(z_i(\theta_0))-{\cal{F}}_1(z_i(\theta_1))|,
\end{equation}
\begin{equation}
  {\cal{W}}_0 ={{\cal{W}}_{||}+{\cal{W}}_\perp \over 2} ={nB\sigma_0 R \over 2} 
  \sum_{i=1}^\infty \Phi(Gi)Gi |{\cal{F}}_2(z_i(\theta_0))-{\cal{F}}_2(z_i(\theta_1))| +R|\theta_0 -\theta_1|n\sigma_A,
\end{equation}  
where
\begin{equation}
  {\cal{F}}_1(z)= -{2\over 3}\sqrt{1-z}-{(36z+4) \over 45}(1-z)^{3/2}-{2\over 15}(1-z)^{5/2}+{z^3\over 3}\ln{1+\sqrt{1-z}\over 1-\sqrt{1-z}},
\end{equation}
\begin{equation}
  {\cal{F}}_2(z)= -{8\over 3}\sqrt{1-z}+({47\over 45}+{2\over 5}z)(1-z)^{3/2} +(1-z)^{5/2} +(z+{z^2 \over 2} -{z^3 \over 6})\ln{1+\sqrt{1-z}\over 1-\sqrt{1-z}}.
\end{equation}
The variable $z_i$ depends on the entrance angle $\theta_0$ and on the final angle $\theta_1$ accordingly
to Eq.~(7). Besides, the functions ${\cal{F}}_1,\,{\cal{F}}_2$ are =0 when $|z| > 1$.    These relations are valid for the case where $\theta_0$ and $\theta_1$ have the same sign.
In the case where these values have a different sign, the correct result is the sum of the two values,
with the first value equal to the probability from $\theta_0$ till $0$ and the second value equal to the 
probabilty from $0$ till $\theta_1$.   

It should be noted that this description is valid for a thin enough crystal where $W_{||},\,W_{\perp} \ll 1$.
For a thicker crystal, a more complicated theory should be used (see for example \cite{MV1}) which takes into
account the changing of the polarization states during the propagation of the photons. 
The corresponding differential probabilities can be found in literature \cite{TM,BKS}.

\section{Examples of calculation}
The method has been applied to compute the expected results with two different 
photon energies: 120 and 1000~GeV. 
The first value has been chosen given the experimental possibility offered by the INSURAD test setup on the 
CERN SPS H4 beamline, while the second value represents approximately the limit of applicability of the method. \\
Fig.~\ref{fig-2} illustrates the behaviour of the total probability of pair production as a function of the
entrance angle $\theta_0$. One can see the non-symmetrical curves with respect to the entrance angle.
At both energies the region of orientation where the difference ${\cal{W}}_{||}-{\cal{W}}_\perp$ is a measurable 
quantity is evident.

The differential spectra of the created positrons (electrons) are shown in Fig.~\ref{fig-3} for two different 
$\theta_0$ entrance angles. For positive angles smaller than the angle $\theta_0$ corresponding
to a maximum of the $d{\cal{W}}_{||}/dx - d{\cal{W}}_\perp /dx$ value, 
the spectra have a bell like shape (see Fig.~\ref{fig-3}a), while for angles larger than this angle, the typical 
form is presented in Fig.~\ref{fig-3}b. 

The spectra in Fig.~\ref{fig-3} are valid for particles which were created in the bulk of the crystal (at the 
moment of creation). After their creation, these particles (electrons and positrons) can lose their energy 
due to coherent radiation processes in the electric fields of the crystal. However, 
these processes are relevant only for small angles
with respect to the tangency direction \cite{CKMY}. Hence, one can expect that the particles created at large
enough $\theta_0$ angles have a large probability to conserve their initial energy till their exit from the crystal.      

Fig.~\ref{fig-4} illustrates the behavior of the ${\cal{W}}_0$ probability and the $A={\cal{W}}_0/{\cal{D}}_0$ 
asymmetry for different bending radii of the silicon single crystal (for a (110) planar orientation). 

\section{A possible setup for the experimental validation}
To measure the pair photoproduction probability in a bent crystal, a high energy 
$\gamma$ ray beam is needed which can be created via bremsstrahlung 
in an amorphous material (Rt, amorphous radiator) as shown in Fig.~\ref{fig-5} which presents a schematic 
of the possible setup. \\
In this proposed setup, an electron beam of 120~GeV is considered; such a beam has 
already been characterized in the INSURAD test at the CERN SPS H4 beamline where electron and
positron beams with energies up to 200~GeV are available.
The primary electron crosses a pair of silicon microstrip detectors (S1-S2) which determine 
its trajectory and therefore the approximate direction of the photons which can be generated 
during the radiator (Rt) crossing. \\
A bending magnet placed after the radiator deflects the primary electron beam to avoid the 
production of other photons in the interaction of the electron beam with the crystal; considering a magnet 
of 3~Tm, the primary electron minimum displacement at the end of the magnet is $\simeq$1~cm, leaving enough space to 
place the bent crystal whose transversal dimension is usually of the order of some millimeters. 
The bending magnet has the additional task to sweep away the electron-positron pair that can be generated 
inside the radiator and would constitute a background in the measurement. 

The photons produced in the target cross the magnet and impinge on the crystal (C), where an 
electron-positron pair can be photoproducted. After the crystal a silicon microstrip detector 
identifies the pair presence and its position before it crosses a second smaller magnet 
(M2, $\simeq$0.5~Tm). M2 separates the trajectory of the two particles of the pair 
as a function of their energy so that the S4 silicon microstrip detectors (acting as a spectrometer 
together with M2) can compute their energy. The S4 dimensions and position determine the 
measurable energy range; a 20~cm detector at 0.5~m from the magnet should allow to measure the 
secondary particles in the energy range from 1 to 60~GeV, thus covering the majority of the 
high energy photoconverted  $\gamma$ quanta.

After the magnets (at around 10~m) the primary electron and the photons which didn't convert in the crystal have 
different directions and can be detected by two different calorimeters, C$\gamma$ and Ce. 
The C$\gamma$ calorimeter measures the energy of the non converted photons while the energy 
of the primary electron can be reconstructed with the spectrometer method, using the position 
information provided by the S5 silicon microstrip detector. The Ce calorimeter would be used 
only for beam monitoring, to measure the purity of the primary beam since it allows to distinguish 
electrons from muons and negative hadrons, that can be present if the beam is a secondary one, as is the case for 
the SPS leptonic extracted beamline.

In principle, the setup in Fig.~\ref{fig-5} should allow to reconstruct the event 
chain which follows a trigger produced by a primary incoming electron. The detectors measure the 
energy of the charged and neutral secondary particles as well as the energy fraction 
remaining in the primary electron. In this way it is possible to check, on an event by event basis, 
that the total measured energy corresponds to the initial one.

The setup shown in Fig.~\ref{fig-5} allows also the pre-alignment of the crystal, 
which is performed removing the radiator target and switching the magnets off. In this configuration, 
the two silicon microstrip detectors (S1-S2) are able to measure the incoming angle (before the crystal) 
of the particle 
while S3 and S4 measure its outgoing angle after the crystal to compute the crystal deflection angle. 
Since the crystal is mounted on a goniometer, it can be rotated until the channeling and volume reflection 
effects appear allowing to compute the misalignment between the beam and the crystal itself.

For what concerns the $\gamma$ beam characteristics, two cases can be considered:
\begin{itemize}
  \item \textit{thin radiator}: the production of more than one photon per incoming electron 
    is improbable, thus the gamma beam intensity and spectral distribution can be measured during the 
    crystal data taking given the fact that the photon energy is either measured using the pair (when 
    conversion has happened) or the C$\gamma$ calorimeter (no conversion)
 \item \textit{thick radiator}: statistically more than one photon per incoming electron 
   is produced, but only the total irradiated energy is measured; the gamma beam intensity and 
   spectral distribution should be inferred through a montecarlo simulation. The advantage of this 
   second configuration is a higher statistics.
\end{itemize}
Finally, the photoproduction probability dependence on the photons linear polarization can be investigated
substituting the amorphous radiator with a straight crystal oriented in coherent bremsstrahlung.
This kind of radiator provides a photon beam with a high degree of linear polarization, ranging from
30\% to 75\% depending on the photon spectral region which is taken into account \cite{TM, BKS, Apy}.
The linear polarization of the emitted photons is parallel to the electric field of the radiator crystal
which in turn depends on its orientation. Therefore, using a goniometer with two degrees of rotational 
freedom to orient the radiator crystal, it is possible to change the photons linear polarization from 
a prevalence of W$_{||}$ (the radiator crystal and the bent one are oriented in the same way, 
Fig.~\ref{fig-6}a) to W$_\perp$ (the crystals are oriented perpendicularly one with respect to the other, 
Fig.~\ref{fig-6}b).
The comparison between the photoproduction probability measured using the photon beam with a prevalence
of W$_{||}$ photons and W$_\perp$ ones allows to measure the photoproduction dependence on the polarization.

\section{Conclusions}

This paper shows how high energy photons moving in bent single crystals with large enough bending radii 
can meet the local conditions for coherent photoproduction. These conditions are violated only in a small range  
when the direction of the photon motion is very close to the tangent direction. According to our estimations 
(for a (110) silicon plane, $E_\gamma \le 1000$~GeV and $R>2$~m), 
this area is about $\pm 20~\mu$rad. For photon energies $\le$1000~GeV, the pair production process is far 
from these angles.
However, with the increase of the photon energy the first harmonic tends to zero and moving towards the TeV range, 
the character of the photoproduction process will change \cite{BT,FM}. \\
We think that these considerations may be extended to the axial case of bent single crystals.

It should be noted that the considered pair production process in bent crystals analyzed in this paper is similar 
to the process of radiation emission of over-barrier positrons (electrons) experimentally investigated in \cite{WS1,AA}. 
The comparison with the theoretical predictions in that case was satisfactory. Theory \cite{CKMY} predicts correctly the 
range and form of the radiation energy losses spectra 
but with an intensity 1.7 (1.2) larger than what experimentally measured for positrons (electrons). 
The first calculations were performed on the basis of the quasi classical description which was developed in 
\cite{BKS}. Then in \cite{MV2} it was demonstrated
that for not large energies the theory of coherent bremsstrahlung in a local longitudinal area of the bent crystal 
could be used to describe radiation emission (see Eq.~(13)). 
A similar approach was also proposed in \cite{AKS, MB} but without taking into consideration the condition of 
volume reflection which accounts for the difference in the soft part of the photon spectrum. 
However, the results presented in all these papers (for large enough bending radii and
not large particle energies) demonstrate the validity of the coherent bremsstrahlung theory in a local area
of the bent crystal. Considering the similarity between coherent pair production and bremsstrahlung, the 
description presented in this paper can be accounted for \footnote{Recently new evidence for this fact has 
been obtained 
in the INSURAD test at the H4 beamline with a very good agreement between simulation and observations 
(see \cite{MP}).}.   
%\pagebreak[4]	

The pair  photoproduction process in bent single crystals may be used to obtain a linearly polarized
high energy $\gamma$-beam and to determine the degree of linear polarization of such beams.
In principle, there is the possibility of transforming the linear polarization into a circular one \cite{MV3,Apy},
but this requires further investigation. The study of this field is interesting due to the new opportunity of 
searching axion-like particles (see for example \cite{AR} and literature herein), offered by the unique nature of 
birefringence for light and high energy photons (the electromagnetic vacuum polarization). \\
Another possible field of application of the photoproduction process in bent single crystals is 
the cleaning of the gamma part of the halos of future electron positron colliders. 
In fact, one may expect the electromagnetic cascade in an oriented bent crystal
to be significantly shorter than in an amorphous medium. The use of a multicrystal system could be 
foreseen for this goal \cite{WS2}.

It should be also noted that the probability and asymmetry functions presented in this paper are smoother and broader 
in angle for bent crystals compared to straight ones ($R$ tends to $\infty$). For this reason bent crystals are more 
convenient for the applications described above. 

The paper presents also a possible experimental setup to perform the measurements, on the basis of 
what has already been done for the INSURAD experiment on the CERN SPS extracted beamlines.
 
\section{Acknowledgments}
This work was supported by the RFBR Grant No.08-02-01453-a and by the joint Grant RFBR 09-02-92431-KE-a
and EINSTEIN Consortium (Italy). One of the authors (V.~A.~M) would like
to thank the Cariplo Foundation Fellowship for the support and the opportunity for joint work at the 
Insubria University.

\newpage
\section{Figure captions}
{Fig.1. Production of $e^\pm$-pairs by a 
        high energy photon moving in a bent single crystal. ($xyz$) is the local Cartesian 
        coordinate system. The $y$-axis
        is directed perpendicularly to the plane of the figure; $d$ is the interplanar 
        distance, $\theta_0$ is the
        entrance angle and $\theta$ is the variable angle relative to the (110)-plane of the single crystal.  }

{Fig.2. Production of $e^\pm$-pairs by a 
        high energy photon moving in a bent single crystal. ($xyz$) is the local Cartesian 
        coordinate system. The $y$-axis
        is directed perpendicularly to the plane of the figure; $d$ is the interplanar 
        distance, $\theta_0$ is the
        entrance angle and $\theta$ is the variable angle relative to the (110)-plane of the single crystal.  }        
 
\begin{figure}
\scalebox{0.7}{\includegraphics{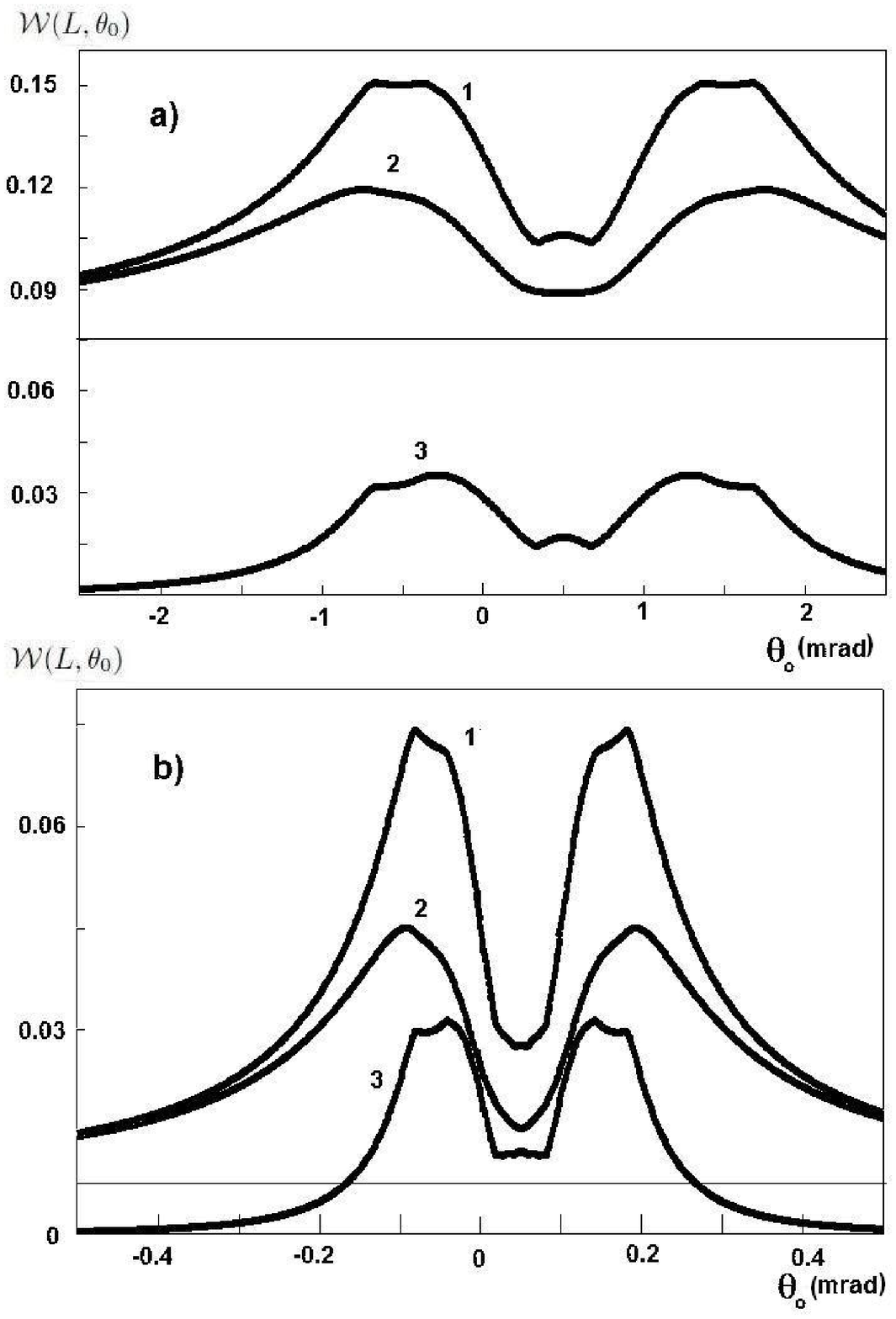}}
{\caption{
              }
\label{fig-2}}
\end{figure}

\begin{figure}
\scalebox{0.7}{\includegraphics{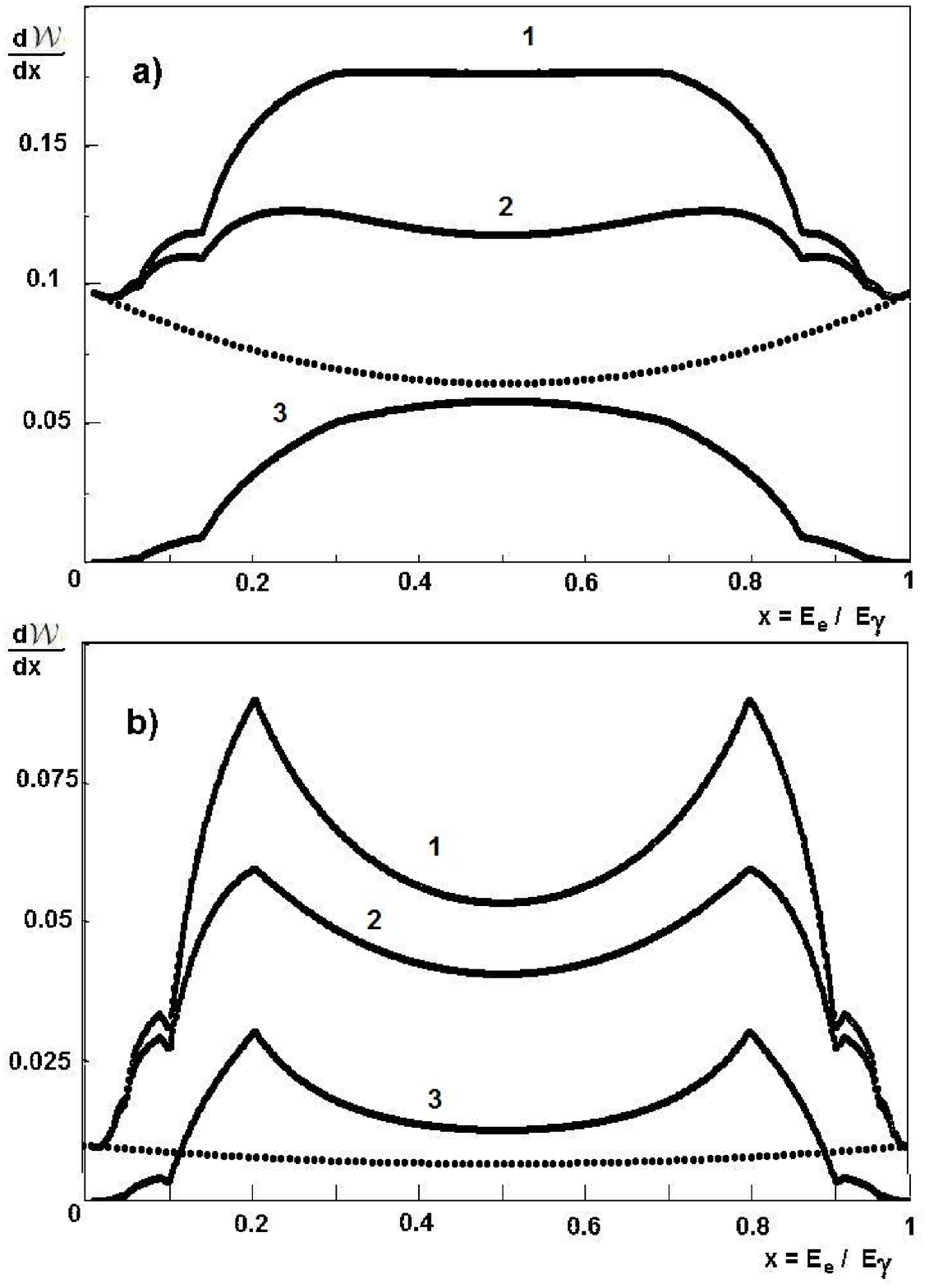}}
{\caption{
              }
\label{fig-3}}
\end{figure}

\begin{figure}
\scalebox{0.7}{\includegraphics{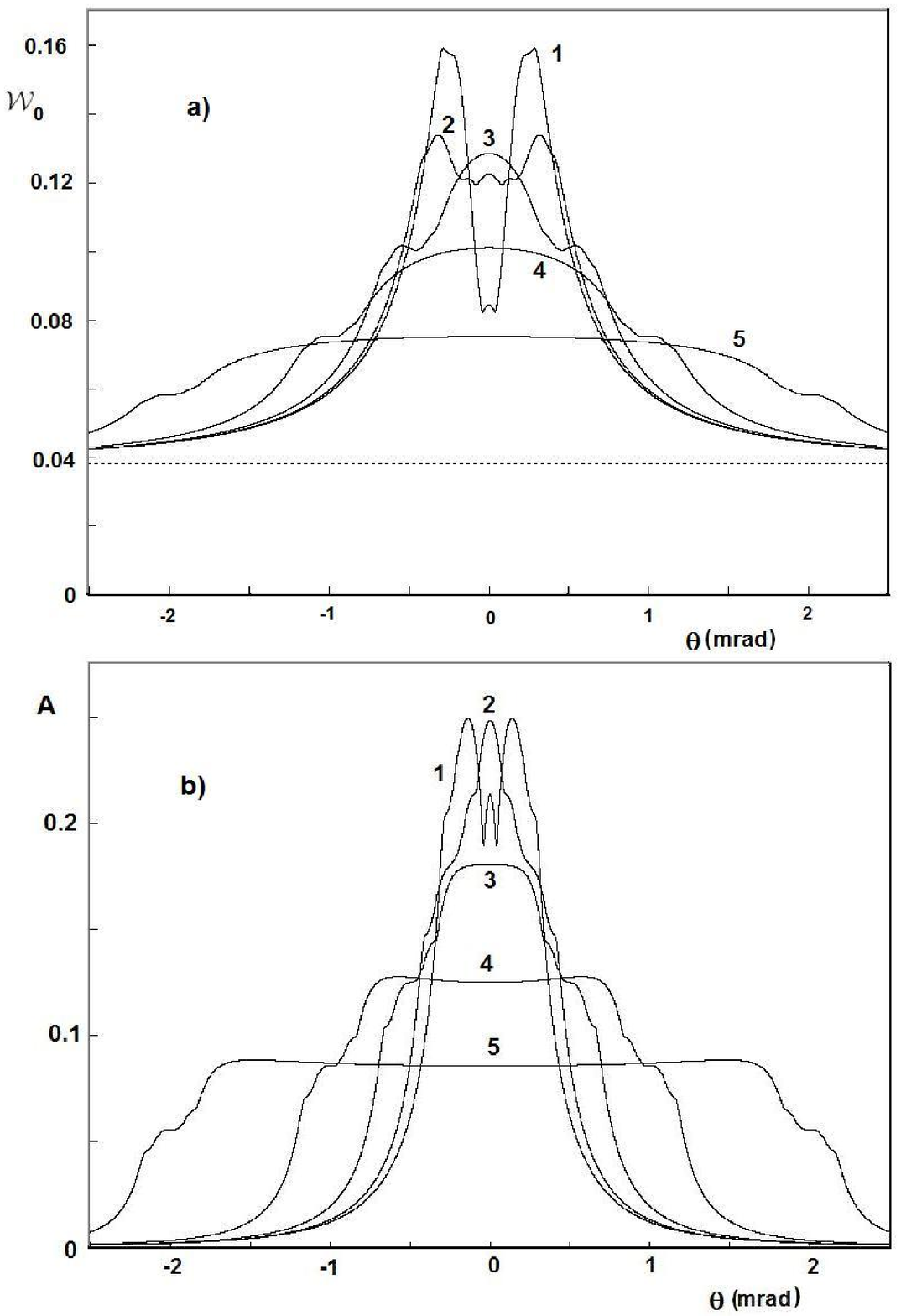}}
{\caption{
              }
\label{fig-4}}
\end{figure}

\begin{figure}
\scalebox{0.27}{\includegraphics{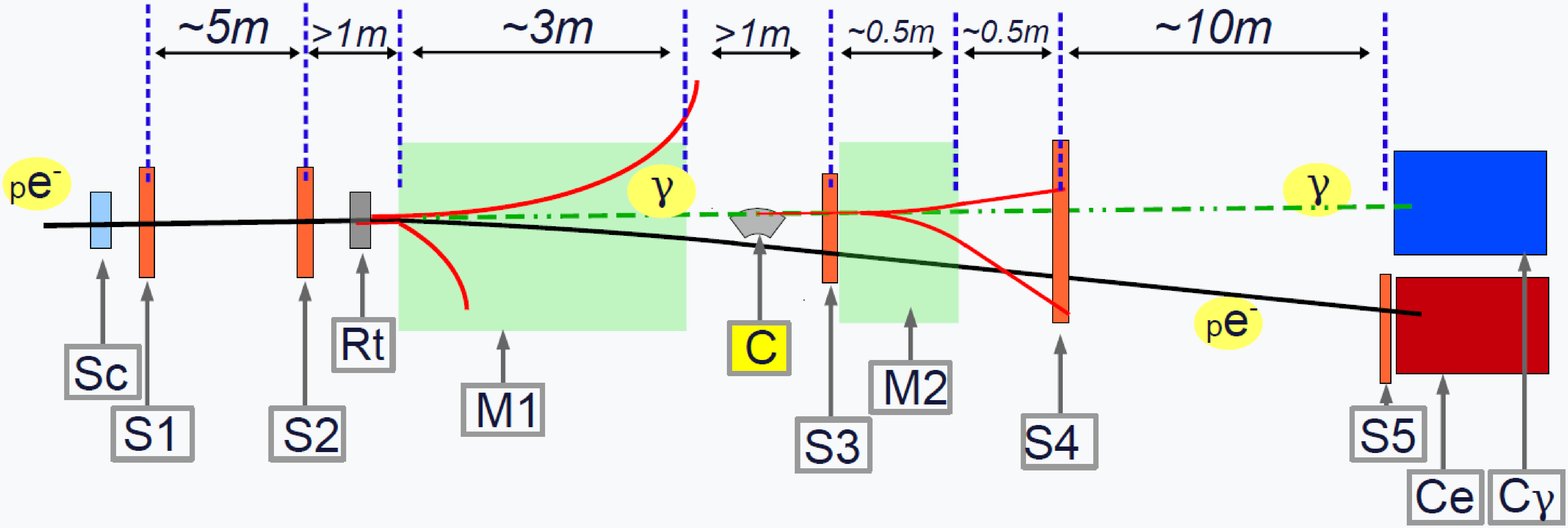}}
{\caption{
              }
\label{fig-5}}
\end{figure}

\begin{figure}
\scalebox{0.27}{\includegraphics{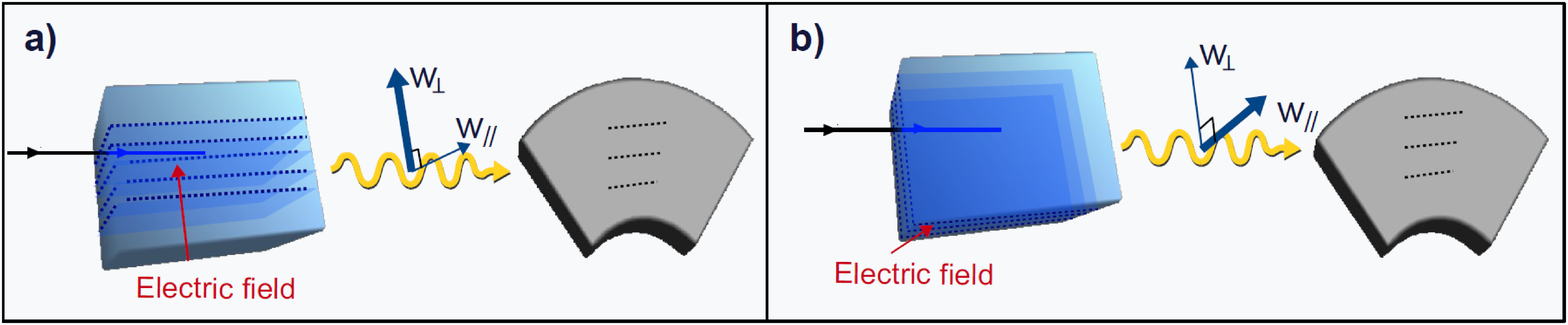}}
{\caption{
              }
\label{fig-6}}
\end{figure}

{Fig. 3. Spectral distributions of the electrons (positrons) produced in a 
        single bent silicon crystal at the conditions of Fig.~\ref{fig-2}, for the two cases (a) and (b). 
        The entrance angle is $\theta_0$= +1.5~mrad (a) and +0.2~mrad (b). The numbers near the curves indicate 
        the probabilities as in Fig.~\ref{fig-2}.   
        The black point curves are the ones for nonoriented single crystals.}
  
{Fig. 4. The probability ${\cal{W}}_0$ (a) and asymmetry $A$ (b) for a 0.5~cm single silicon crystal  
        at different bending radii
        as a functions of the orientation angle $\theta$. Curves 1-5 correspond to $R= 20, 10, 5, 2.5$ and $1.25$~m 
        ($\theta=0$ for a symmetric orientation of the single crystal). The point curves correspond to a non oriented crystal.  
        The photon energy is equal to 500~GeV.}

{Fig. 5. Schematic of a possible setup to measure the pair production yield in a bent crystal 
      as a function of its orientation. It contains the following elements: Sc, scintillator for the trigger; 
      S1-S5, silicon microstrip detectors to track the charged particles; Rt, Radiator target to produce the 
      photon beam; M1, magnetic field ($\simeq$1~T$\times$3~m) to sweep the primary beam and the unwanted 
      pair produced in the amorphous target; C, bent crystal mounted on a goniometer; M2, 
      magnetic field ($\simeq$1~T$\times$0.5~m) to measure the produced pair energy; Ce, electromagnetic 
      calorimeter to tag the electron (the primary beam purity could be smaller than 100\%); C$\gamma$, 
      electromagnetic calorimeter to measure the energy of the photons that do not convert in the crystal.}
 
{ Fig. 6. To produce a polarized photon beam the amorphous radiator can be substituted with a straight 
      crystal oriented in coherent bremsstrahlung. a) Both the radiator crystal and the bent one are 
      oriented in the same way, that is forming a small angle with respect to the rotation axis perpendicular to the 
      figure; b) the radiator crystal orientation is perpendicular with respect to the previous one, that is  
      the crystalline planes are directed horizontally in the radiator crystal and vertically in the amorphous one.}       
\clearpage
\newpage 

\newpage 
\newpage 
\newpage 
\newpage 

\newpage 
\end{document}